\documentclass[12pt]{PNASTWO}
\usepackage[margin=22mm]{geometry}
\usepackage{graphicx}
\newcommand{\equn}[1]{\begin{equation}\label{#1}}
\newcommand{\eqan}[1]{\begin{eqnarray}\label{#1}}
\newcommand{\eqa}{\begin{eqnarray}}
\newcommand{\equ}{\begin{equation}}
\newcommand{\nuqe}{\end{equation}}
\newcommand{\uqe}{\end{equation}}
\newcommand{\naqe}{\end{eqnarray}}
\newcommand{\aqe}{\end{eqnarray}}
\newcommand{\nonu}{\nonumber}

\newcommand{\la}{\langle}
\newcommand{\ra}{\rangle}
\newcommand{\e}{{\rm e}}
\providecommand{\binom}[2]{{#1\choose#2}}
\begin{document}
\title{Analytical distributions for stochastic gene expression}
\author{Vahid Shahrezaei\affil{1}{Centre for Non-linear Dynamics, Dept.\ of Physiology, McGill University, 3655 Promenade Sir William Osler, Montreal, Quebec H3G 1Y6, Canada}
\and
Peter S.\ Swain\thanks{To whom correspondence should be addressed: Email: swain@cnd.mcgill.ca, Tel: +1 514 398 4360, Fax: +1 514 398 7452.}
\affil{1}{}
}
\maketitle

\begin{article}
\begin{abstract}

Gene expression is significantly stochastic making modeling of genetic networks challenging. We present an approximation that allows the calculation of not only the mean and variance but also the distribution of protein numbers. We assume that proteins decay substantially slower than their mRNA and confirm that many genes satisfy this relation using high-throughput data from budding yeast. For a two-stage model of gene expression, with transcription and translation as first-order reactions, we calculate the protein distribution for all times greater than several mRNA lifetimes and thus qualitatively predict the distribution of times for protein levels to first cross an arbitrary threshold. If in addition the promoter fluctuates between inactive and active states, we can find the steady-state protein distribution, which can be bimodal if promoter fluctuations are slow. We show that our assumptions imply that protein synthesis occurs in geometrically distributed bursts and allows mRNA to be eliminated from a master equation description. In general, we find that protein distributions are asymmetric and may be poorly characterized by their mean and variance. Through maximum likelihood methods, our expressions should therefore allow more quantitative comparisons with experimental data. More generally, we introduce a technique to derive a simpler, effective dynamics for a stochastic system by eliminating a fast variable.

\end{abstract}


\dropcap{G}ene expression in both prokaryotes and eukaryotes is inherently stochastic \cite{ozbudak02,elowitz02,blake,raser}. This stochasticity is both controlled and exploited by cells, and, as such, must be included in models of genetic networks \cite{kaern05,shahrezaeiRev}. Here we will focus on describing intrinsic fluctuations, those generated by the random timing of individual chemical reactions, but extrinsic fluctuations are equally important and arise from the interactions of the system of interest with other stochastic systems in the cell or its environment \cite{swain02,shahrezaei}. Typically, experimental data are compared with predictions of mean behaviors and sometimes with the predicted standard deviation around this mean because protein distributions are often difficult to derive analytically, even for models with only intrinsic fluctuations. 

We will propose a general, although approximate, method for solving the master equation for models of gene expression. Our approach exploits the difference in lifetimes of mRNA and protein and is valid when the protein lifetime is greater than the mRNA lifetime. Typically, proteins exist for at least several mRNA lifetimes, and protein fluctuations are determined by only time-averaged properties of mRNA fluctuations. Following others \cite{swain02,thattai01,kepler,swain04}, we will use this time-averaging to simplify the mathematical description of stochastic gene expression.

For many organisms, single cell experiments have shown that gene expression can be described by a three-stage model \cite{blake,raser,golding,chubb,raj}. The promoter of the gene of interest can transition between two states \cite{kepler,ko,peccoud,karmakar}, one active and one inactive. Such transitions could be from changes in chromatin structure, from binding and unbinding of proteins involved in transcription \cite{blake,raser,golding}, or from pausing by RNA polymerase \cite{voliotis}.  Transcription can only occur if the promoter region is active. Both transcription and translation, as well as the degradation of mRNAs and proteins, are usually modelled as first-order chemical reactions \cite{kaern05}.

By taking the limit of a large ratio of protein to mRNA lifetimes, we will study the three-stage model and a simpler two-stage version where the promoter is always active. For this two-stage model, we will derive the protein distribution as a function of time. We will derive the steady-state protein distribution for the full, three-stage model. We also include expressions for the corresponding mRNA distributions \cite{raj,peccoud} in the Supporting information.

\subsection*{A two-stage model of gene expression.}
\begin{figure}[ht]
\centering
\includegraphics[width=160mm]{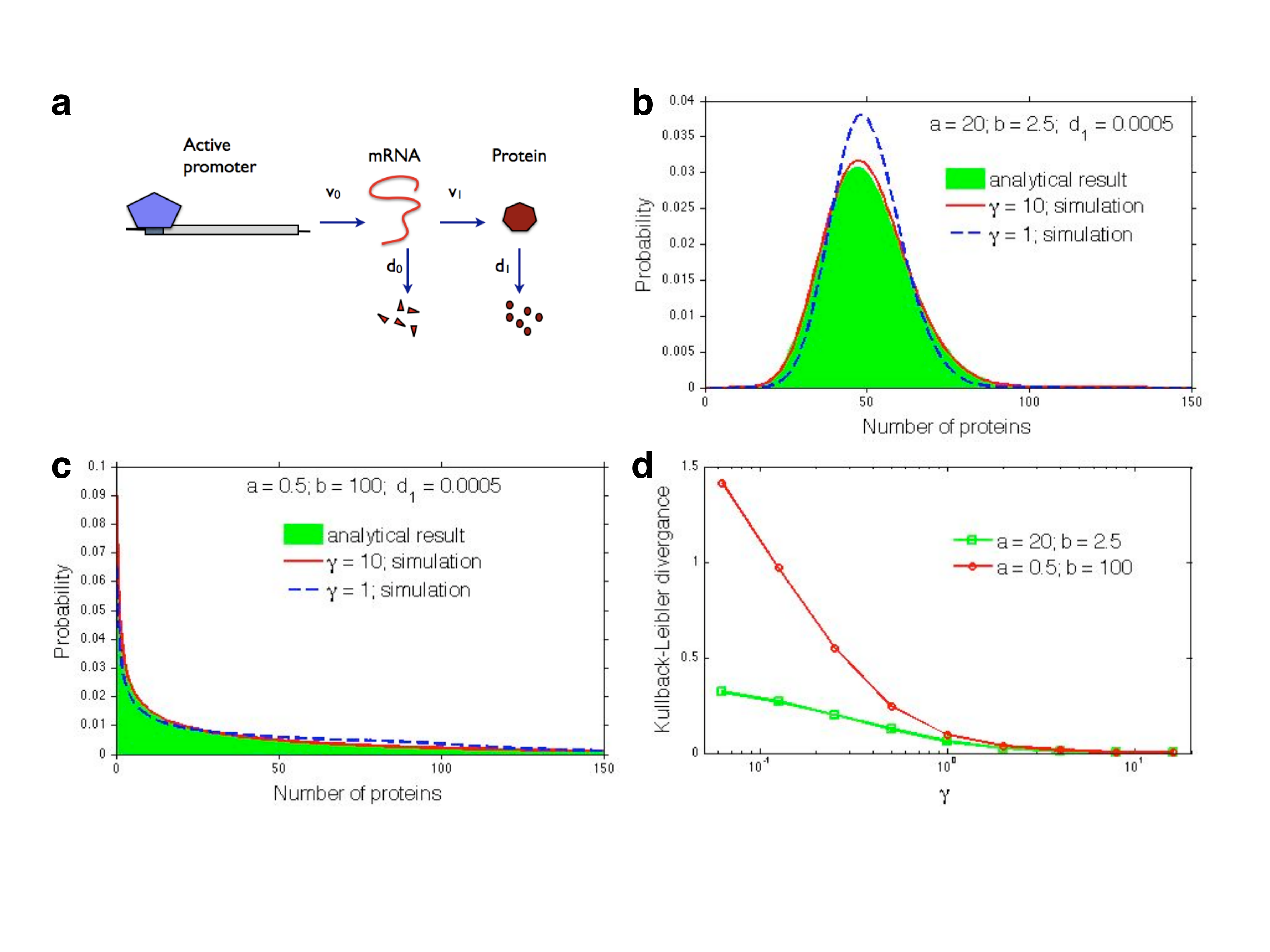}
\caption{Predictions and simulations for a two-stage model of gene expression. {\bf a} Both transcription and translation are modelled as first-order processes: transcription occurs with a probability $v_0$ per unit time and translation with a probability of $v_1$ per unit time. Degradation of mRNA and protein are also both first-order processes: mRNA degrades with a probability $d_0$ per unit time and protein degrades with a probability $d_1$ per unit time. {\bf b} and {\bf c} A comparison of Eq.\ \ref{pn0}, shown as the distribution in green, and stochastic simulations for large and small $\gamma$. Protein distributions can be either peaked or have a maximum only at $n=0$ \cite{friedman}. The mean number of mRNAs, $a/\gamma$, is either 2 or 20 in {\bf b} and either 0.05 or 0.5 in {\bf c}. {\bf d} The accuracy of Eq.\ \ref{pn0} improves with larger $\gamma$. The Kullback-Leibler divergence between the analytical and simulated protein distributions is plotted as a function of $\gamma$. For $\gamma$ greater than 1, the distributions become almost indistinguishable.}
\label{fig1}
\end{figure}

We will first consider the model of gene expression in Fig.\ \ref{fig1}a \cite{thattai01}. This model assumes the promoter is always active and so has two stochastic variables: the number of mRNAs and the number of proteins. The probability of having $m$ mRNAs and $n$ proteins at time $t$ satisfies a master equation:
\eqan{ma}
\frac{\partial P_{m,n}}{\partial t} &=& v_0 (P_{m-1,n} - P_{m,n}) + v_1 m (P_{m,n-1}-P_{m,n}) \nonu \\
& & + d_0 \Bigl[ (m+1) P_{m+1,n}- m P_{m,n} \Bigr] \nonu \\
& & + d_1 \Bigl[ (n+1)P_{m,n+1} - n P_{m,n} \Bigr]
\naqe
with $v_0$ being the probability per unit time of transcription, $v_1$ being the probability per unit time of translation, $d_0$ being the probability per unit time of degradation of an mRNA, and $d_1$ being the probability per unit time of degradation of a protein. By defining the generating function, $F(z',z)$, by $F(z',z) = \sum_{m,n} z'^m z^n P_{m,n}$, we can convert Eq.\ \ref{ma} into a first-order partial differential equation:
\equn{ma2}
\frac{\partial F}{\partial v} - \gamma \left[ b(1+u) - \frac{u}{v} \right] \frac{\partial F}{\partial u} + \frac{1}{v} \frac{\partial F}{\partial \tau} = a \frac{u}{v} F ,
\nuqe
where we have rescaled \cite{friedman}, with $a= v_0/d_1$, $b=v_1/d_0$, $\gamma= d_0/d_1$, and $\tau= d_1 t$, and where $u=z'-1$ and $v= z-1$.

If the protein lifetime is much greater than the mRNA lifetime and $\gamma \gg 1$, Eq.\ \ref{ma2} can be solved using the method of characteristics. Let $r$ measure the distance along a characteristic which starts at $\tau=0$ with $u=u_0$ and $v=v_0$ for some constant $u_0$ and $v_0$, then Eq.\ \ref{ma2} is equivalent to \cite{zwillinger}
\equn{ma3}
\begin{array}{lclclcl}
\frac{dv}{dr} = 1 &;& \frac{d\tau}{dr} = \frac{1}{v}\\ 
\gamma^{-1} \frac{du}{dr} = \frac{u}{v}-b(1+u)  &;& \frac{dF}{dr} = \frac{a u}{v} F .
\end{array}
\nuqe
Consequently direct integration implies $r=v=v_0 \e^\tau$. For $\gamma \gg 1$,  $u(v)$ obeys (Supporting information)
\equ
u(v) \simeq \left( u_0 - \frac{b v_0}{1-b v_0}\right) \e^{-\gamma b (v-v_0)} \left( \frac{v}{v_0} \right)^\gamma + \frac{bv}{1-bv}
\uqe
or
\equ
u(v) \simeq \frac{bv}{1-bv}
\uqe
as $v = v_0 \e^\tau > v_0$ for $\tau > 0$. When $\gamma \gg 1$, $u$ rapidly tends to a fixed function of $v$: for most of a protein's lifetime, the dynamics of mRNA is at steady-state. The generating function then obeys
\equn{ma4}
\frac{dF}{dv} \simeq \frac{a b}{1-b v} F .
\nuqe
Intuitively, Eq.\ \ref{ma4} arises from Eq.\ \ref{ma2} because large $\gamma$ causes the term in square brackets in Eq.\ \ref{ma2} to tend to zero to keep $F(u,v)$ finite and well defined. Eq.\ \ref{ma4} describes only the distribution for protein numbers: $F(u,v)$ is just a function of $v$. Terms of higher order in $\gamma^{-1}$ will depend on $u$. Large $\gamma$ implies that most of the mass of the joint probability distribution  of mRNA and protein is peaked at $m= 0$: $P_{m,n} \simeq P_{0,n}$. 

\begin{figure}[ht]
\centering
\includegraphics[width=160mm]{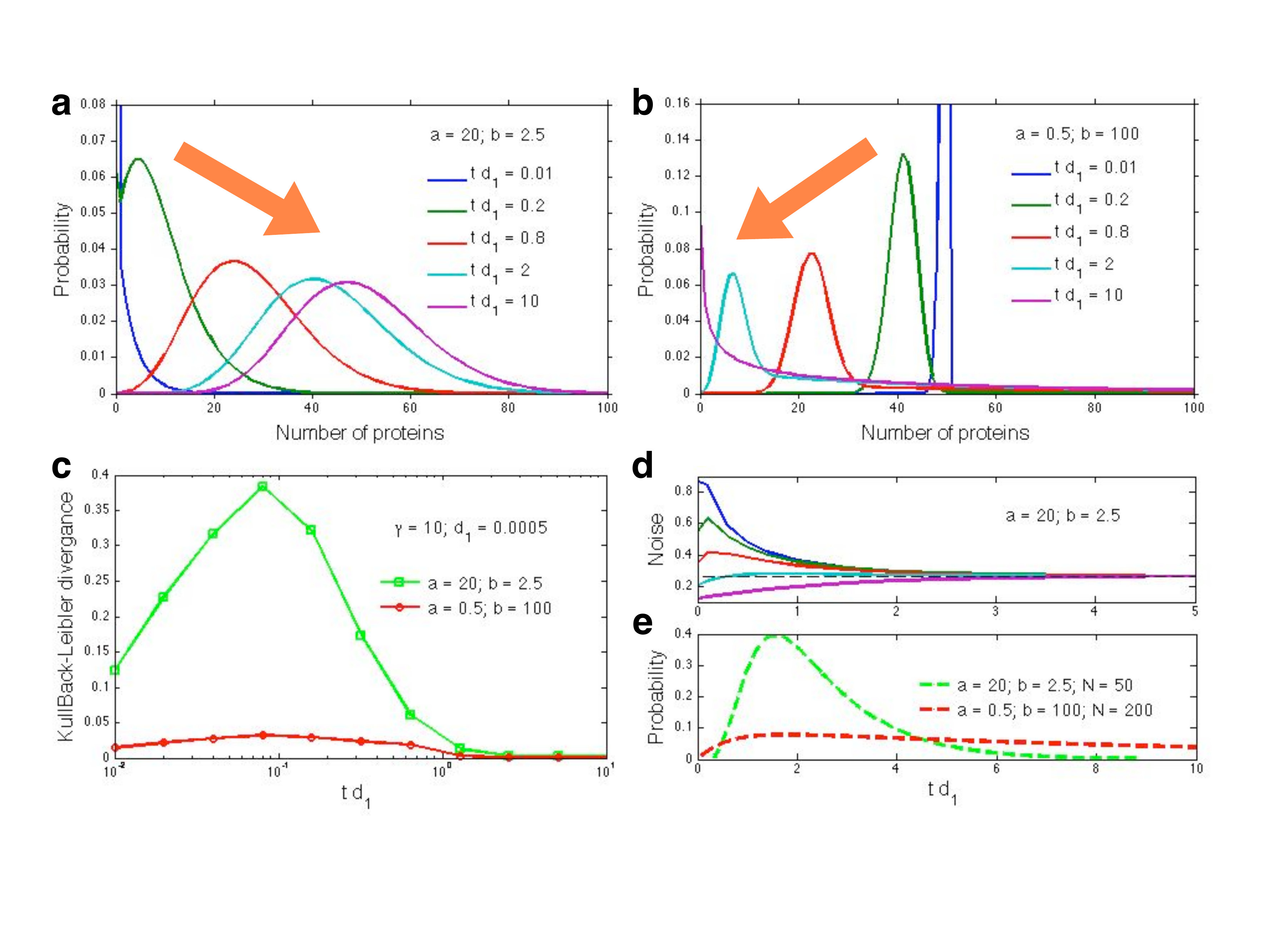}
\caption{Predictions for the time-dependent solution of the two-stage model of gene expression. {\bf a, b} The distribution of protein numbers at different times with time increasing in the direction of the arrow. Parameters in {\bf a} correspond to Fig.\ \ref{fig1}b. There are zero proteins initially. Parameters in {\bf b} correspond to Fig.\ \ref{fig1}c. There are 50 proteins initially. {\bf c} The Kullback-Leibler divergence for the distributions of {\bf a} and {\bf b}. The divergence decreases as $\tau= t d_1$ grows above $\gamma^{-1}$. It is small for small times because both the simulations and the calculations start from the same initial distribution. {\bf d} Noise in protein numbers as a function of time. Initially, proteins have a negative binomial distribution chosen to have a particular magnitude of noise. The noise at steady-state is shown by a dashed line. {\bf e} The calculated distributions for the first time protein levels reach a given threshold, $N$, if initially there are zero proteins. These distributions are qualitative with the probability typically underestimated for small $t d_1$. They obey a renewal equation \cite{vankampen}, which we solve numerically.
}
\label{fig2}
\end{figure}

We can find the probability distribution for protein numbers as a function of time by integrating Eq.\ \ref{ma4}. Integration gives
\equn{ft}
F(z,\tau)= \left[ \frac{1-b(z-1)\e^{-\tau}}{1+b-bz} \right]^a
\nuqe
assuming that no proteins exist at $\tau= 0$. From the definition of a generating function, expanding $F(z)$ in $z$ gives (Supporting information)
\eqan{pt}
P_n(\tau) &=& \frac{\Gamma(a+n)}{\Gamma(n+1)\Gamma(a)} \left( \frac{b}{1+b} \right)^n \left( \frac{1+b \e^{-\tau}}{1+b} \right)^a \nonu \\
&& \times \, _2F_1 \left( -n, -a, 1-a-n ; \frac{1+b}{\e^\tau +b} \right) 
\naqe
where $P_n(\tau)= P_{0,n}(\tau)$. Here $_2F_1(a,b,c; z)$ is a hypergeometric function and $\Gamma$ denotes the gamma function \cite{abramowitz}. Eq.\ \ref{pt} is valid when $\gamma \gg 1$, $\tau \gg \gamma^{-1}$ to allow the mRNA distribution to reach steady-state, and $a$ and $b$ are finite. The mean, $\la n \ra = ab (1-\e^{-\tau})$, and the variance, $\la n^2 \ra - \la n \ra^2 = \la n \ra (1 + b + b\e^{-\tau})$, of Eq.\ \ref{pt} agree with earlier results \cite{thattai01}. At steady-state $\tau \gg 1$ and 
\equn{pn0}
P_n= \frac{\Gamma(a+n)}{\Gamma(n+1) \Gamma(a)} \left( \frac{b}{1+b} \right)^n \left( 1 - \frac{b}{1+b} \right)^a 
\nuqe
which is a negative binomial distribution. We verified Eq.\ \ref{pt} and Eq.\ \ref{pn0} with stochastic simulations using the Gibson-Bruck version \cite{gibson} of the Gillespie algorithm \cite{gillespie} and the {\it Facile} network complier and stochastic simulator \cite{siso}. If $\gamma \gg 1$, Eq.\ \ref{pn0} accurately predicts the distribution described by Eq.\ \ref{ma} (Fig.\ \ref{fig1}b and \ref{fig1}c), but it fails as expected for smaller $\gamma$. This effect can be quantified by calculating the Kullback-Leibler divergence between the predicted and simulated distributions for different $\gamma$ (Fig.\ \ref{fig1}d). Eq.\ \ref{pt} is illustrated in Fig.\ \ref{fig2}a and \ref{fig2}b. As well as $\gamma \gg 1$, times with $\tau > \gamma^{-1}$ are necessary for negligible Kullback-Leibler divergences (Fig.\ \ref{fig2}c).

Eq.\ \ref{pt} allows complete characterization of the Markov process underlying the two-stage model. The `propagator' probability, $P_{n|k}(\tau)$, which is the probability of having $n$ proteins at time $\tau$ given $k$ proteins initially, satisfies (Supporting information)
\equn{ptk}
P_{n|k}(\tau) = \sum_{r=0}^k  \binom{k}{r} P_{n-r}(\tau) \left( 1 - \e^{-\tau} \right)^{k-r} \e^{-r \tau} 
\nuqe
where $P_n(\tau)=0$ if $n<0$. With Eqs.\ \ref{pt} and \ref{ptk}, two-stage gene expression is in principle completely characterized for $\gamma \gg 1$ and $\tau \gg \gamma^{-1}$. For example, we can calculate how the noise in protein numbers, $\eta$ (their standard deviation divided by their mean), changes with time. If protein numbers initially have a distribution $P_k^{(0)}$, then at a time $\tau$ their distribution will be $\sum_k P_{n|k}(\tau) P_k^{(0)}$. The noise of this distribution can either increase, decrease, or behave non-monotonically as time increases (Fig.\ \ref{fig2}d). We can also calculate non-steady state auto-correlation functions and first-passage time distributions for protein levels to first cross a threshold, $N$ (with some standard numerics). In general, such distributions are only qualitative because contributions from times with $\tau < \gamma^{-1}$ are always relevant. Accuracy can be improved by having $\gamma \gg 10$  and a sufficiently high threshold (Fig.\ \ref{fig2}e and Supporting information).

We can derive Eq.\ \ref{pn0} more intuitively. An mRNA undergoes a competition between translation and degradation because ribosomes and degradosomes bind to it mutually exclusively \cite{mcadams97}. For each competition, the probability of a ribosome binding to the mRNA is $\frac{v_1}{v_1+d_0}= \frac{b}{1+b}$. If we assume that proteins have longer lifetimes than mRNAs ($\gamma \gg 1$), then each protein synthesized from a given mRNA will not on average be degraded before the mRNA is degraded. On protein timescales, all the proteins synthesized from an mRNA will appear to be synthesized simultaneously (Fig.\ 5 in Supporting information). Consequently, the probability of $r$ new proteins being produced by the synthesis and degradation of one mRNA is equal to the probability of an mRNA being translated $r$ times. This probability is \cite{mcadams97}
\equn{geo}
P_r= \left( \frac{b}{1+b} \right)^r \left( 1 - \frac{b}{1+b} \right) 
\nuqe
which is a geometric, or `burst', distribution. Alternatively, we can consider the lifetime $t'$ of each mRNA. This lifetime is stochastic and satisfies $P(t') = d_0 \e^{-d_0 t'}$, the distribution expected for any first-order decay process \cite{vankampen}. Proteins synthesis is also first order, and the number of proteins, $r$, synthesized by an mRNA during its lifetime satisfies a Poisson process: $\frac{(v_1 t')^r}{r!} \e^{-v_1 t'}$ \cite{vankampen}. Consequently, the probable number of proteins synthesized from a particular mRNA is given by 
\equn{anon}
P(r) = \int_0^\infty dt' \, d_0 \e^{-d_0 t'} \frac{(v_1 t')^r}{r!} \e^{-v_1 t'} 
\nuqe
which integrates to Eq.\ \ref{geo}.  Eq.\ \ref{geo} is equivalent to an exponential distribution with a parameter $\lambda$ where $\lambda= -\log\left( 1 - \frac{b}{1+b} \right)$  \cite{prochaska}. If $b < 1$, then $\lambda \simeq b$. Exponential bursts of protein synthesis have been characterized experimentally \cite{cai,yu}. We note that Eq.\ \ref{geo} has a generating function $f(z)= (1+b-b z)^{-1}$. 

Given that the synthesis and degradation of one mRNA generates a burst of $r$ proteins, then the number of proteins at steady-state is given by the typical number of mRNAs synthesized during a protein lifetime,  $\frac{v_0}{d_1}= a$, and the $r_i$ for each mRNA. The number of proteins $n$ will be sum of these $r_i$. If we assume that there are sufficient ribosomes and charged tRNAs, then translation from each mRNA is independent. The generating function of a sum of independent variables is the product of their individual generating functions \cite{vankampen}. Consequently, the generating function for $P_n$, $F(z)$, satisfies
\equn{voila}
F(z)= \prod_{i=1}^a f(z) = (1 + b - b z)^{-a}
\nuqe
which is Eq.\ \ref{ft} when $\tau \gg 1$, and so derives Eq.\ \ref{pn0}.

By assuming explicitly that protein synthesis occurs in bursts, we can derive an effective master equation for gene expression that considers only proteins, but implicitly includes mRNA fluctuations \cite{friedman,paulsson00}. We will show that this master equation has Eq.\ \ref{pt} as its solution and so is equivalent to the large $\gamma$ approximation to Eq.\ \ref{ma}, the master equation for both mRNA and protein. If we assume that each mRNA synthesized leaves behind a burst of $r$ proteins then
\eqan{mab}
\frac{\partial P_n}{\partial \tau} &=& a \left[ \left( 1-\frac{b}{1+b} \right) \sum_{r=0}^n \left(\frac{b}{1+b} \right)^r P_{n-r} - P_n \right] \nonu \\
& & + (n+1) P_{n+1} - n P_n 
\naqe
where the size of each burst has been determined by Eq.\ \ref{geo} \cite{paulsson00}. Eq.\ \ref{mab} has Eq.\ \ref{ft} as its generating function (Supporting information). By introducing bursts of protein synthesis, mRNA fluctuations can be absorbed into a one-variable master equation provided $\gamma \gg 1$. Friedman {\it et al.} used a continuous version of this approach with an exponential burst distribution inspired by their experimental results \cite{cai,yu}. They derived a gamma distribution for steady-state protein numbers \cite{friedman}. Eq.\ \ref{pn0} tends to this distribution 
\equn{pn1}
P_n \rightarrow \frac{n^{a-1} \e^{-n/b}}{b^a \Gamma(a)} 
\nuqe
for large $n$ (Supporting information). Friedman {\it et al.} also demonstrated that the burst approximation remains valid when negative or positive feedback is included \cite{friedman}.

In summary, we have shown that exploiting the difference between protein and mRNA lifetimes through a large value of $\gamma$, but finite $a$ and $b$, allows powerful mathematical simplifications. Large $\gamma$ implies that mRNA is at steady-state for most of the lifetime of a protein and that the probability mass of the joint distribution of protein and mRNA is peaked at zero mRNAs, although the mean number of mRNAs need not be zero (Fig. \ref{fig1}). The number of proteins translated from an mRNA obeys a geometric distribution in both the two-stage and three-stage models \cite{mcadams97}, but large $\gamma$ implies that the proteins translated from an mRNA all appear, on protein timescales, simultaneously so that the synthesis and degradation of an mRNA leaves behind a geometric burst of proteins. If $\gamma < 1$, then proteins synthesized from a particular mRNA will be degraded as further proteins are synthesized, and the distribution describing the number of proteins remaining once the mRNA is degraded will no longer be geometric. Explicitly including geometric bursts accurately describes the effects of mRNA fluctuations on the distribution of protein numbers when $\gamma \gg 1$. It allows the model of Fig.\ \ref{fig1}a to be described by a one-variable master equation: Eq.\ \ref{mab}.

\subsection*{A three-stage model of gene expression.}

\begin{figure}[ht]
\centering
\includegraphics[width=160mm]{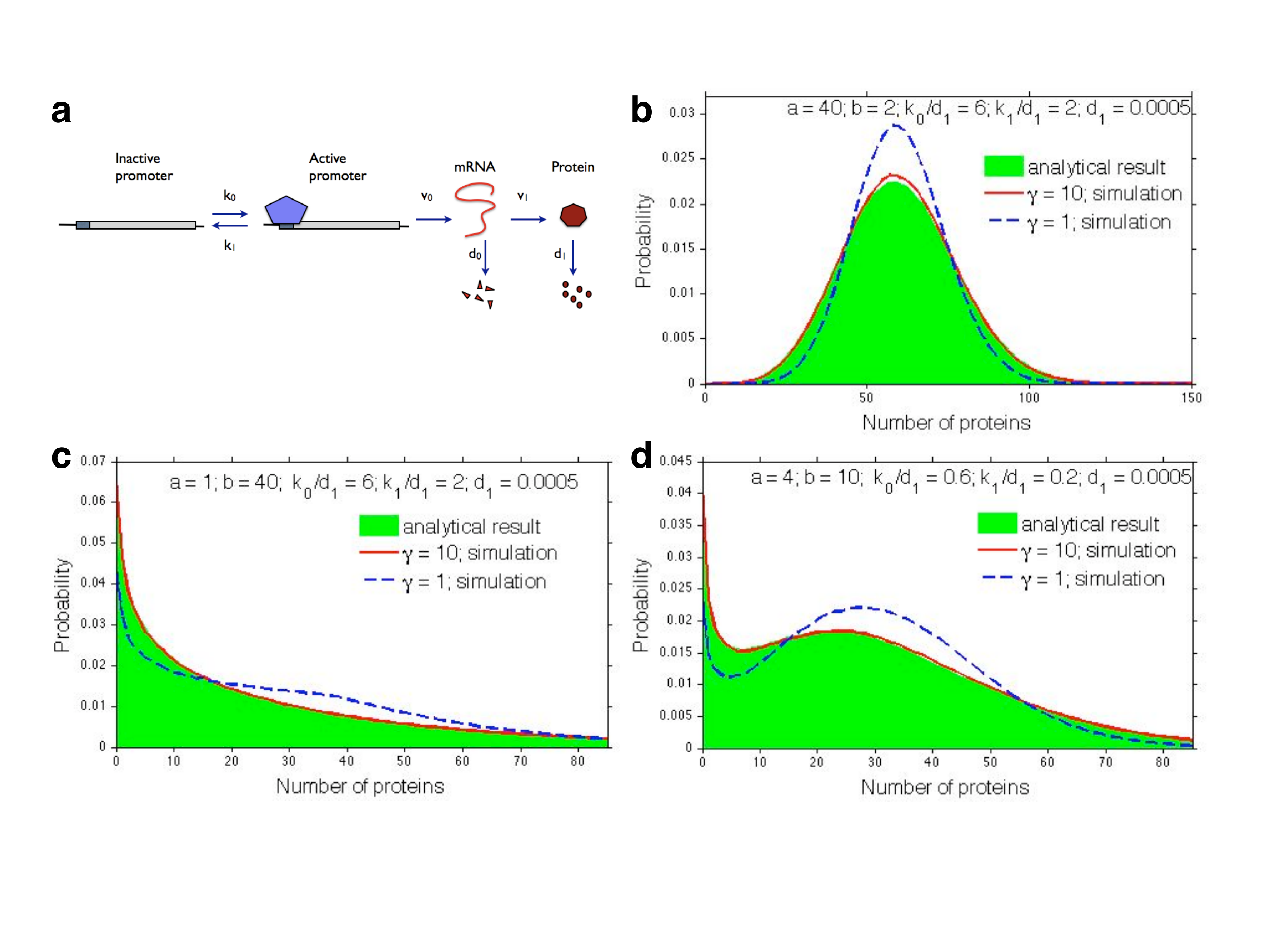}
\caption{Predictions and simulations for a three-stage model of gene expression.  {\bf a} The region of the DNA containing the promoter region transitions between inactive and active forms with probabilities per unit time of $k_0$ and $k_1$. As an example, we show the TATA-box binding protein driving the transition. {\bf b}, {\bf c}, and {\bf d} A comparison of Eq.\ \ref{pn}, shown as the distribution in green, and stochastic simulations for large and small $\gamma$. The mean number of mRNAs, $\frac{a k_0}{\gamma(k_0+k_1)}$, is either 3 or 30 in {\bf b}, 0.075 or 0.75 in {\bf c}, and either 0.3 or 3 in {\bf d}. }
\label{fig3}
\end{figure}

We next consider the full three-stage model of gene expression (Fig.\ \ref{fig3}a). We find the protein distribution for this system by taking the large $\gamma$ limit of the master equation. Let $P^{(0)}_{m,n}$ be the probability of having $m$ mRNAs and $n$ proteins when the DNA is inactive and  $P^{(1)}_{m,n}$ be the probability of having $m$ mRNAs and $n$ proteins when the DNA is active. We then have two coupled equations:
\eqan{macouple}
\frac{\partial P_{n,m}^{(0)}}{\partial \tau} &=&  \kappa_1 P_{m,n}^{(1)} - \kappa_0 P_{m,n}^{(0)} + (n+1) P_{m,n+1}^{(0)} - n P_{m,n}^{(0)} \nonu \\
& & + \gamma \Bigl[ (m+1) P_{m+1,n}^{(0)} - m P_{m,n}^{(0)} \nonu \\
& & + b m \left( P_{m,n-1}^{(0)} - P_{m,n}^{(0)} \right) \Bigr] \\
\frac{\partial P_{n,m}^{(1)}}{\partial \tau} &=&  -\kappa_1 P_{m,n}^{(1)} + \kappa_0 P_{m,n}^{(0)} + (n+1) P_{m,n+1}^{(1)} - n P_{m,n}^{(1)} \nonu \\
& & + a \left( P_{m-1,n}^{(1)} - P_{m,n}^{(1)} \right) \nonu \\
& &  +  \gamma \Bigl[ (m+1) P_{m+1,n}^{(1)} - m P_{m,n}^{(1)} \nonu \\
& & + b m \left( P_{m,n-1}^{(1)} - P_{m,n}^{(1)} \right) \Bigr] \label{macouple0}
\naqe
where $\kappa_0= k_0/d_1$ and $\kappa_1= k_1/d_1$. 

We solve Eqs.\ \ref{macouple} and \ref{macouple0} at steady-state by taking the large $\gamma$ limit of the equivalent equations for their generating functions (a generating function is defined for each state of the promoter). Our approach is a natural extension of the method used to solve the two-stage model (Supporting information). We find that
\eqan{pn}
P_n &=&  \frac{\Gamma(\alpha +n)\Gamma(\beta + n)\Gamma(\kappa_0+\kappa_1)}{\Gamma(n+1)\Gamma(\alpha)\Gamma(\beta)\Gamma(\kappa_0+\kappa_1+n)} \nonu \\
&& \times \left( \frac{b}{1+b} \right)^n \left( 1- \frac{b}{1+b} \right)^\alpha  \nonu \\
&& \times \,  _2F_1\left(\alpha +n, \kappa_0+\kappa_1-\beta, \kappa_0+\kappa_1+n; \frac{b}{1+b} \right)  \nonu \\
&&
\naqe
where
\eqan{al}
\alpha & =& \frac{1}{2} \left( a + \kappa_0+\kappa_1 + \phi \right) \\
\beta &=& \frac{1}{2} \left( a + \kappa_0+\kappa_1 - \phi \right)  \label{be}
\naqe
and $\phi^2=(a+\kappa_0+\kappa_1)^2-4 a \kappa_0$. Eq.\ \ref{pn} is valid when $\gamma \gg 1$ and $a$ and $b$ are finite. The mean of this distribution is $\la n \ra = \frac{a b k_0}{k_0+k_1}$ and the protein noise, $\eta$, satisfies
\equn{noise}
\eta^2 = \frac{1}{\la n \ra} + \gamma^{-1} \frac{1}{\la m \ra} + \frac{d_1}{d_1 + k_0+k_1} \eta^2_{D}
\nuqe
where $\la m \ra$ is the mean number of mRNAs, and is inversely proportional to $\gamma$, and $\eta_D$ is the noise in the active state of DNA: $\eta_D^2=k_1/k_0$ \cite{raser}. As well as a Poisson-like term expected for any birth-and-death process, protein noise has time-averaged contributions from fluctuations in the number of mRNAs and fluctuations in the state of DNA. We verify Eq.\ \ref{pn} by simulation in Fig.\ \ref{fig3}. 

The protein distribution for the three-stage model can have similar behavior to the two-stage model of Fig.\ \ref{fig1}a, but it can also generate a bimodal distribution with a peak both at zero and non-zero numbers of molecules (Fig.\ \ref{fig3}d). This bimodality is not a reflection of an underlying bistability, but arises from slow transitions driving the DNA between active and inactive states \cite{kaern05,karmakar,hornos,pirone}.

As expected, Eq.\ \ref{pn} recovers the negative binomial distribution under certain conditions. It tends to Eq.\ \ref{pn0} when $\kappa_1 \rightarrow 0$: the DNA is then always active at steady-state. When $\kappa_1 = 0$, Eqs.\ \ref{al} and \ref{be} imply that $\alpha = a$ and $\beta = \kappa_0$, and recall that $_2F_1(a,0,c;z) = 1$ for all $a$, $c$, and $z$. Similarly, when $\kappa_0$ and $\kappa_1$ are both large, but $\kappa_0/\kappa_1$ is fixed, then $\alpha \rightarrow \kappa_0 + \kappa_1$ and $\beta \rightarrow \frac{\kappa_0 a}{\kappa_0+\kappa_1}$. Consequently,
\equn{lim}
P_n \rightarrow \frac{\Gamma(\beta+n)}{\Gamma(n+1) \Gamma(\beta)} \left( \frac{b}{1+b} \right)^n \left( 1 - \frac{b}{1+b} \right)^\beta
\nuqe
because $_2F_1(a,b,a;z)= (1-z)^{-b}$. With fast switching of the DNA between active and inactive states, Eq. \ref{pn} becomes Eq.\ \ref{pn0}, but with $a$ replaced by $ \frac{\kappa_0 a}{\kappa_0+\kappa_1}$.

\subsection*{Discussion.}

\begin{figure}[ht]
\centering
\includegraphics[width=150mm]{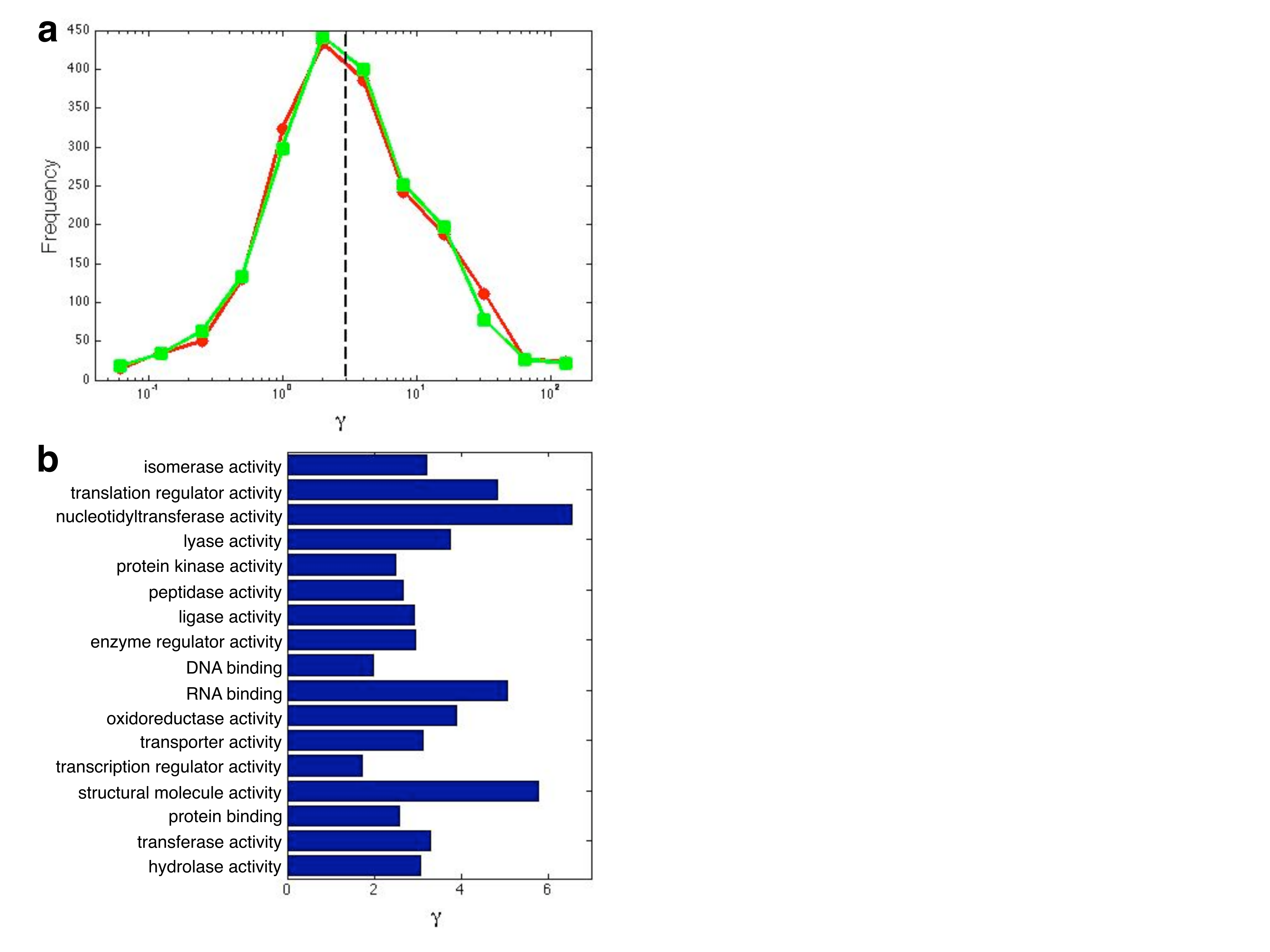}
\caption{The ratio of the protein to mRNA lifetime, $\gamma$, for 1,962 genes in budding yeast. {\bf a} Most proteins have $\gamma > 1$. Protein lifetimes are from Belle {\it et al.} \cite{belle} and mRNA lifetimes are from Grigull {\it et al.} (circles) \cite{grigull} or from Wang {\it et al.} (squares) \cite{wang}. The median of $\gamma$ is $ \simeq 3$ (shown by a dashed line), while its mean is greater than 10 (although this value is probably erroneously high because of outliers). Overall, we found little correlation between mRNA and protein lifetimes. {\bf b} The median value of $\gamma$ for genes in different gene ontology classes. We plot the mean of the medians for the two datasets. Errors in the medians are approximately 25\% (using 1000 bootstrap samples for each gene ontology class). Gene annotations are from the {\it Saccharomyces cerevisiae} genome database ({\tt www.yeastgeonome.org}). }
\label{fig4}
\end{figure}

We have shown we can calculate distributions for protein numbers by assuming protein lifetimes are longer than mRNA lifetimes while $a$, the number of mRNAs transcribed during a protein lifetime, and $b$, the number of proteins translated during a mRNA lifetime, are finite. Fig.\ \ref{fig4}a shows the ratio $\gamma$ measured for almost 2,000 genes in budding yeast. Around 80\% of the genes have $\gamma$ greater than one and the median value is approximately 3 (we include the data set in Supporting information). We therefore expect our predicted distributions to be widely applicable in budding yeast. In bacteria, too, $\gamma$ is expected to be greater than 1 because mRNA lifetimes are usually minutes (they are typically tens of minutes in yeast) and protein lifetimes are often determined by the length of the cell cycle (typically 30 or more minutes) \cite{bremer}. 

Values of $\gamma > 1$ reduce protein fluctuations by allowing more averaging of the underlying mRNA fluctuations (Eq.\ \ref{noise}). We indeed observe a small, but statistically significant, negative correlation between total noise and $\gamma$ using the data of Newman {\it et al.} \cite{newman} (a rank correlation of $\simeq -0.2$ with a P value of 10$^{-6}$). In Fig.\ \ref{fig4}b, we have calculated the median $\gamma$ for yeast genes in different gene ontology classes. All classes have a median $\gamma > 1$. Proteins involved in transferring nucleotidyl groups, which include RNA and DNA polymerases, have high median $\gamma > 5$, presumably because high stochasticity in these proteins can undermine many cellular processes. Similarly, proteins that contribute to the structural integrity of protein complexes have a median $\gamma > 5$. Large fluctuations can vastly reduce the efficiency of complex assembly by preventing complete complexes forming because of a shortage of one or more components \cite{swain04,fraser}. Perhaps surprisingly transcription factors have a low median $\gamma > 1$. Although low $\gamma$ does increase stochasticity, it can allow quick response times if the protein degradation rate is high. A high protein degradation rate may also keep numbers of transcription factors low to reduce deleterious non-specific chromosomal binding. 

We show that protein synthesis occurs in bursts in both the two- and the three-stage model when $\gamma \gg 1$. Such bursts of gene expression have been measured in bacteria and eukaryotes \cite{golding,chubb,raj,cai,yu}. They allow mRNA to be replaced in the master equation by a geometric distribution for protein synthesis for all times greater than several mRNA lifetimes if their source is translation and the protein lifetime is substantially longer than the mRNA lifetime. Such an approach has already been proposed \cite{friedman}, but without determining its validity. Similarly, if mRNA fluctuations are negligible, the master equation reduces to one variable (protein), and describing the protein distribution becomes substantially easier \cite{hornos,walczak}. 

An important problem in systems biology is to determine which properties of biochemical networks and the intracellular environment must be modeled to make accurate, quantitative predictions. As well as obscuring the process driving the observed phenotype, models more complex than needed are harder to correctly parameterize and to simulate to generate predictions. Our results show that complexity, here two states of the promoter, can be modelled by effective parameters that under certain conditions will give accurate predictions of the entire distribution of protein numbers: Eq.\ \ref{lim}. Alternatively, they show that not only the mean and variance \cite{pedraza08} but also the protein distribution may not have enough information to determine the biochemical mechanism generating gene expression from measurements of protein levels: Eqs.\ \ref{pn0} and \ref{lim}. Such effects are likely to be compounded by non-steady-state dynamics (Fig.\ \ref{fig2}) and extrinsic fluctuations. Collecting data on the corresponding mRNA distribution may disfavor the two-stage over the three-stage model because mRNA distributions in the three-stage model can have two peaks even though the protein distribution has only one \cite{shahrezaeiRev}. In general, though, time series measurements, preferably with and without perturbations, may provide the most discriminative power \cite{pedraza08}.

Experimental measurements are best compared with the predicted distribution rather than its mean, standard deviation, or mode. Both the protein and mRNA distributions are typically not symmetric and may not be unimodal. Consequently, the mean and the mode can be significantly different, and the standard deviation can be a poor measure of the width of the distribution at half maximum \cite{samoilov}. Such distributions are poorly characterized by the commonly used coefficient of variation because they are not locally Gaussian around their mean (Figs.\ \ref{fig1}--\ref{fig3}). In addition, fitting moments to find model parameters can be challenging. Moments, more so than distributions, are functions of combinations of parameters and can also be badly estimated without large amounts of data, particularly for asymmetric distributions. We therefore believe a Bayesian or maximum likelihood approach is most suitable where the experimental protocol is replicated by the fitting procedure and explicitly accounts for the shape of the distribution and the number of measurements. For example, irrespective of how many measurements are available, the likelihood of the data for a particular set of parameters can always be determined from the assumed distribution of protein numbers. Our analytical expressions will greatly speed-up such approaches by avoiding large numbers of simulations and by aiding in deconvolving extrinsic fluctuations which can substantially change the shape of protein distributions \cite{shahrezaei}. 

Our results should also allow more general fluctuation analyzes of gene expression data. Such analyzes convert fluorescence measurements into absolute units (numbers of molecules) by exploiting that the magnitude of fluctuations is determined by the number of molecules independently of how those numbers are measured \cite{rosenfeld06}. Converting into absolute units is essential if information from different experiments is to be combined into a larger, predictive framework, a goal of systems biology.

More generally, our approach is an example of a technique to simplify the dynamics of a stochastic system by exploiting differences in timescales. We remove a fast stochastic variable through replacing a constant parameter (the parameter $a$) by a time-dependent parameter (the burst distribution) whose variation captures the effects of fluctuations in the fast variable on the dynamics of the slow one \cite{shibata}.

\subsection*{Acknowledgments.}
P.S.S.\ holds a Tier II Canada Research Chair. V.S.\ and P.S.S.\ are supported by N.S.E.R.C.\ (Canada). We would like to thank an anonymous referee for showing us Eq.\ \ref{anon}.



\newpage

\newpage

\newpage

\newpage

\end{article}
\end{document}